\documentclass[aps,prd,reprint,superscriptaddress,floatfix,showpacs]{revtex4-2}
\usepackage{graphicx}
\usepackage{amsmath}
\usepackage[capitalise]{cleveref }
\usepackage[english]{babel}
\usepackage{placeins}
\usepackage{longtable}
\usepackage{dcolumn}
\usepackage{xcolor}
\usepackage{xspace} 
\usepackage{xfp}

\usepackage{lineno}
\usepackage[
  separate-uncertainty=true,
  per-mode=symbol-or-fraction,
]{siunitx}
\DeclareSIUnit\evperc{\eV\per\clight}
\DeclareSIUnit\clight{\text{\ensuremath{c}}}

\newcommand{\decay}[2]{\ensuremath{#1\!\to #2}\xspace}

\newcommand{\KS}{\ensuremath{K_S^0}}
\newcommand{\jpsi}{\ensuremath{J/\psi}}

\newcommand{\antiproton}{\ensuremath{\bar p}}
\newcommand{\proton}{\ensuremath{p}}
\newcommand{\pip}{\ensuremath{\pi^+}}
\newcommand{\pim}{\ensuremath{\pi^-}}

\newcommand{\piz}{\ensuremath{\pi^0}}
\newcommand{\BF}{\ensuremath{\mathcal{B}} }

\newcommand*{\ppbareta}{\ensuremath{\decay{\jpsi}{\proton\antiproton\eta}}}
\newcommand*{\twogg}{\ensuremath{\decay{\eta}{\gamma\gamma}}}
\newcommand*{\threepi}{\ensuremath{\decay{\eta}{\pip\pim\piz}}}
\newcommand*{\pizero}{\ensuremath{\decay{\piz}{\gamma\gamma}}}

\newcommand{\SigDecay}{\ensuremath{\ppbareta}}

\def\Ppsi{\ensuremath{\psi}\xspace}
\newcommand{\etapr}{\ensuremath{\eta^{\prime}}\xspace}
\def\Pe {\ensuremath{\mathrm{e}}\xspace}
\def\en {{\ensuremath{\Pe^-}}\xspace}
\def\ep {{\ensuremath{\Pe^+}}\xspace}

\newcommand{\twoggnSig}{259.85}
\newcommand{\twoggnSigErr}{17}

\newcommand{\threepinSig}{58.33}
\newcommand{\threepinSigErr}{10}

\newcommand{\twoggvalue}{1.482}
\newcommand{\twoggstat}{0.001}
\newcommand{\twoggsys}{0.024}

\newcommand{\threepivalue}{1.550}
\newcommand{\threepistat}{0.003}
\newcommand{\threepisys}{0.035}

\newcommand{\BFcombi}{1.496}
\newcommand{\Sigmacombi}{0.023}
\newcommand{\SigmacombiStat}{0.001}

\begin{document}

\title{\boldmath Measurement of the branching fraction of the decay $\SigDecay$}

\author{
\begin{small}
   \begin{center}
M.~Ablikim$^{1}$, M.~N.~Achasov$^{4,c}$, P.~Adlarson$^{75}$, O.~Afedulidis$^{3}$, X.~C.~Ai$^{80}$, R.~Aliberti$^{35}$, A.~Amoroso$^{74A,74C}$, Q.~An$^{71,58,a}$, Y.~Bai$^{57}$, O.~Bakina$^{36}$
% ...
\\
\vspace{0.2cm}
(BESIII Collaboration)\\
\vspace{0.2cm} {\it
$^{1}$ Institute of High Energy Physics, Beijing 100049, People's Republic of China\\
% ...
$^{3}$ Bochum Ruhr-University, D-44780 Bochum, Germany\\
$^{4}$ Budker Institute of Nuclear Physics SB RAS (BINP), Novosibirsk 630090, Russia\\
% ...
$^{35}$ Johannes Gutenberg University of Mainz, Johann-Joachim-Becher-Weg 45, D-55099 Mainz, Germany\\
$^{36}$ Joint Institute for Nuclear Research, 141980 Dubna, Moscow region, Russia\\
% ...
$^{57}$ Southeast University, Nanjing 211100, People's Republic of China\\
$^{58}$ State Key Laboratory of Particle Detection and Electronics, Beijing 100049, Hefei 230026, People's Republic of China\\
% ...
$^{71}$ University of Science and Technology of China, Hefei 230026, People's Republic of China\\
% ...
$^{74}$ University of Turin and INFN, (A)University of Turin, I-10125, Turin, Italy; (B)University of Eastern Piedmont, I-15121, Alessandria, Italy; (C)INFN, I-10125, Turin, Italy\\
$^{75}$ Uppsala University, Box 516, SE-75120 Uppsala, Sweden\\
% ...
$^{80}$ Zhengzhou University, Zhengzhou 450001, People's Republic of China\\
\vspace{0.2cm}
$^{a}$ Deceased\\
% ...
$^{c}$ Also at the Novosibirsk State University, Novosibirsk, 630090, Russia\\
% ...
   }\end{center}
\end{small}
}

\begin{abstract}

A high precision measurement of the branching fraction of the decay
$\ppbareta$ is performed using \num{10087(44)e6} $\jpsi$ events
recorded by the {BESIII} detector at the {BEPCII} storage ring. The
branching fractions of the two decays $\decay{\jpsi}{\proton\antiproton\eta(\twogg})$ and 
$\decay{\jpsi}{\proton\antiproton\eta(\threepi})$
are measured individually to be 
$\mathcal{B}(\decay{\jpsi}{\proton\antiproton\eta(\decay{\eta}{\gamma\gamma})})
=
(\twoggvalue\pm\num{\twoggstat}\pm\num{\twoggsys})\times\,10^{-3}$
and
$\mathcal{B}(\decay{\jpsi}{\proton\antiproton\eta(\decay{\eta}{\pip\pim\piz})})
=
(\threepivalue\pm\num{\threepistat}\pm\num{\threepisys})\times\,10^{-3}$,
where the first uncertainties are statistical and the second
systematic. Both results are compatible within their uncorrelated systematic uncertainties. The combined result is
$\BF(\ppbareta)=(\BFcombi\pm\num{\SigmacombiStat}\pm\num{\Sigmacombi})\times\,10^{-3}$
where the first uncertainty is the combined statistical uncertainty
and the second one the combined systematic uncertainty of both
analyses, incorporating correlations between them.
In addition, the $\proton\antiproton$ threshold region is
investigated for a potential threshold enhancement, and no evidence
for one is observed.  

\end{abstract}

\date{\today}


\maketitle

\section{Introduction} The Standard
Model of particle physics describes many aspects in the field of subatomic physics with high
precision. However, in the non-perturbative regime of Quantum Chromodynamics (QCD), many details are still not understood, and not all
experimental observations can be explained. In addition, accurate
predictions for particle interactions, resonance spectra, and decay
processes are difficult to obtain due to the non-Abelian character of
the underlying theory. One example is the spectrum of the $N^*$
states, the excited nucleon resonances. There are many $N^*$
resonances predicted by various theoretical models, e.g. the constituent quark model \cite{Ronniger:2011td} or Lattice QCD calculations \cite{Edwards:2011jj}. 
However, only a few of them have been experimentally confirmed so far. Many N* resonances, listed in the review of the Particle Data Group (PDG) \cite{pdg}, have been reported only by one experiment and need confirmation. In many cases, the properties of the reported resonances are not precisely measured yet.
The huge BESIII data set allows
high precision studies of $\jpsi$ decays e.g. the determination of
branching fractions $\mathcal{B}$ and also the study of the $N^*$
spectrum.

In recent years, several experimental results have been published
about an enhancement near the $\proton\antiproton$ threshold in
radiative charmonium decays $\decay{\jpsi}{\gamma\proton\antiproton}$
and $\decay{\Ppsi'}{\gamma\proton\antiproton}$ \cite{ppbar2003,
ppbar2012}. However, comparable hadronic decays like
$\decay{\jpsi}{X\proton\antiproton}$, where $X$ represents either
$\omega$, $\pi$, or $\eta$, have not shown similar structures
\cite{ppbareta2001, ppbaromega2008, ppbaromega2013,
psiToppbarXCleo}. Other radiative decays into light hadrons like
$\decay{\jpsi}{\gamma\etapr\pip\pim}$ and
$\decay{\jpsi}{\gamma\KS\KS\eta}$ also show structures near the
$\proton\antiproton$ threshold \cite{gammaetaprpipi, X1835-2011,
gammaKSKSeta2015}. Different theoretical interpretations of these
structures have been proposed, such as a $\proton\antiproton$ bound
state with mass $m_X \approx \SI{1.85}{\giga\evperc\squared}$
\cite{ppbar_bound1, ppbar_bound2} or as a glueball, which would
explain the absence of these structures in hadronic decays
\cite{ppbar_glue1, ppbar_glue2}. An overview is given in the review
\cite{threshold_overview}. Since data in the energy range close to the
$\proton\antiproton$ threshold is sparse, these models are not well
constrained by data \cite{ppbar_finalState}. In addition to these
explanations, other effects, such as final state interaction, might
occur in the $\proton\antiproton$ system, which might contribute to 
enhancements near the $\proton\antiproton$ threshold. Therefore, it
is important to search for threshold enhancements with higher
statistics in the decays $\decay{\jpsi}{\proton\antiproton\eta}$ and
$\decay{\jpsi}{\proton\antiproton\piz}$ to better constrain the
models.

In this work, the branching fractions of the decay of
$\decay{\jpsi}{\proton\antiproton\eta}$ with $\twogg$ or $\threepi$
are measured with greatly improved precision in comparison to the
previous measurements. Currently, the world average listed by PDG is
dominated by the measurement taken at BESII, which measured
$\BF(\ppbareta)=(1.91\pm0.02\pm0.17)\times\,10^{-3}$
\cite{oldbes}. The present work improves upon the BESII measurement
with the much larger data set of BESIII, improved analysis techniques
that result in reduced systematic uncertainties, and, crucially, an
improved determination of the global reconstruction efficiency. The
precision is improved by more than a factor of 10. The large number of
events in this final state also allows the exploration of the
threshold region.

\section{BESIII experiment}
The BESIII detector is a magnetic spectrometer~\cite{BESIIISYS:2009}
located at the Beijing Electron Positron Collider
(BEPCII)~\cite{Yu:IPAC2016-TUYA01}. The cylindrical core of the BESIII
detector consists of a helium-based multilayer drift chamber (MDC), a
plastic scintillator time-of-flight system (TOF), and a CsI(Tl)
electromagnetic calorimeter (EMC), which are all enclosed in a
superconducting solenoidal magnet providing a 1.0~T magnetic field
(0.9~T in 2012). The solenoid is supported by an octagonal flux-return
yoke with resistive plate chamber muon-identifier modules interleaved
with steel. The acceptance for charged particles and photons is 93\%
over the $4\pi$ solid angle. The charged-particle momentum resolution
at $1~\si{\giga\eV/\clight}$ is $0.5\%$, and the
specific energy loss ($\mathrm{d}E/\mathrm{d}x$) resolution is $6\%$ for electrons from
Bhabha scattering. The EMC measures photon energies with a resolution
of $2.5\%$ ($5\%$) at $1$~GeV in the barrel (end-cap) region. The time
resolution of the TOF barrel part is 68~ps, while that in the end cap
region was 110~ps. The end cap TOF system was upgraded in 2015 with
multigap resistive plate chamber technology, providing a time resolution of
60~ps, which benefits 87\% of the data used in this analysis~\cite{etof::Guo2019,etof::Li2017}.

\section{\label{sec::DATASETS}Data sets}

In this analysis, the complete $\jpsi$ data set of \num{10087(44)e6}
$\jpsi$ events recorded by the BESIII experiment in the years 2009, 2012, 2018, and 2019
is analyzed. The total number of $\jpsi$ events is
determined using inclusive hadronic $\jpsi$ decays
\cite{nJpsi}. Additionally, the continuum data set at center of mass
(CM) energy $\sqrt{s} = \SI{3.080}{\giga\eV}$ with an overall
luminosity of $168.6\,$pb$^{-1}$ is analyzed to estimate background
contributions from QED processes, beam-gas interactions, and cosmic
rays.  To understand the reconstruction efficiency of the signal
channel as well as the relevant resolutions and limitations of the
detector, Monte Carlo (MC) simulations are used.  The initial $\ep\en$
collision, including initial state radiation, and the generation of
the $\jpsi$ meson are simulated using {\sc kkmc} \cite{ref::kkmc}. The
$\jpsi$ decay and subsequent decays are simulated with the event
generator {\sc evtgen} \cite{ref::evtgen2001, ref::evtgen2008}. Interactions with the detector material are simulated using {\sc
geant4}~\cite{Agostinelli:2002hh}.

Several MC samples are used in this analysis.  Two exclusive samples
of $\num{1e6}$ events were produced to determine the reconstruction
efficiencies of the signal decays $\ppbareta$, with the subsequent
decays of either $\twogg$ or $\threepi$ with $\pizero$.
Since the distributions of the reconstructed data events
deviate from pure phase space (PHSP), these MC samples
are generated using a model obtained with an amplitude analysis, which
will be described in Section \ref{sec:eff}. The decay distribution of
the $\eta$ into three pions follows the $\rm ETA\_DALITZ$ model
\cite{ref::evtgen2001} of {\sc evtgen}, which is adjusted to fit
experimental data.

Additionally, an inclusive MC sample of $\num{10e9}$ $\jpsi$ events is
used to identify possible background contributions.  This sample is
generated to match the number of BESIII $\jpsi$ events and uses a
combination of world average $\BF \text{s}$ from the PDG \cite{pdg} and
effective models from {\sc lundcharm} \cite{ref:lundcharm1,
ref:lundcharm2}.

\section{\label{sec::PID}Event selection} 

The decay $\ppbareta$ is reconstructed using the dominant $\eta$
decays $\twogg$ and $\threepi$, with the $\piz$ subsequently decaying into $\gamma\gamma$.  Consequently, each event is required to contain at least two
photons and two charged tracks in the decay
$\decay{\jpsi}{\proton\antiproton\eta}$ with $\decay{\eta}{\gamma\gamma}$,
or four charged tracks in the decay
$\decay{\jpsi}{\proton\antiproton\eta}$ with $\decay{\eta}{\pip\pim\piz}$.

Charged tracks are required to be reconstructed within the acceptance
of the MDC, satisfying $|\cos\theta|<0.93$, with $\theta$ being the
angle between the reconstructed track and the $z$ axis, which is the
symmetry axis of the MDC.  Additionally, the distance of closest
approach to the interaction point is required to be
$|V_{xy}|<\SI{1}{\centi\meter}$ in the radial direction and
$|V_z|<\SI{10}{\centi\meter}$ along the $z$ axis.  In the $\threepi$
channel, the particle identification (PID) system is used to distinguish protons and charged pions. This system combines measurements of the energy deposited in the MDC~(d$E$/d$x$) and the flight time in the TOF to form likelihoods $\mathcal{L}(h)~(h=p,K,\pi)$ for each hadron $h$ hypothesis.
Protons are identified by imposing the criterion
$\mathcal{L}(p)>\mathcal{L}(\pi)$, while charged pions are identified by requiring
$\mathcal{L}(\pi)>\mathcal{L}(p)$. Since no sizable
kaon background channels could be identified, no requirement on the
kaon likelihood is used. In the $\twogg$ channel, no PID requirement
is applied to the charged tracks, since using the kinematic fit already
suppresses most of the background events.

The photons from $\eta$ and $\piz$ decays are required to have an energy
deposition of more than $\SI{25}{\mega\eV}$ in the barrel part of the
EMC ($|\cos\theta|<\num{0.8}$) and more than $\SI{50}{\mega\eV}$ in
the endcaps of the EMC ($\num{0.86}<|\cos\theta|<\num{0.92}$).  The
angle $\Delta\alpha$ between the photon and nearest charged track should be larger than
$20^\circ$ to exclude bremsstrahlung photons or hadronic split offs
from charged tracks, and especially antiproton interactions within
the calorimeter.  Furthermore, it is required that the EMC shower is
within $\SI{700}{\nano\second}$ after the time of the
collision. Combinations in which both photons are detected in the
endcaps are also rejected, since this improves the overall mass
resolution of the $\eta$ and $\piz$ candidates.

The selected photons are combined into $\eta$ and $\piz$ candidates,
requiring the invariant mass $M_{\gamma\gamma}$ of the two photons to
lie within wide mass windows of $\SI{200}{\mega\eV} \leq
M_{\gamma\gamma} \leq \SI{900}{\mega\evperc\squared}$ for $\eta$ and
$\SI{80}{\mega\evperc\squared} \leq M_{\gamma\gamma} \leq
\SI{180}{\mega\evperc\squared}$ for $\piz$. In the $\threepi$ decay
channel, the invariant mass of the three pions $M_{\pip\pim\piz}$ must
be within the range of $\SI{200}{\mega\evperc\squared} \leq
M_{\pip\pim\piz} \leq \SI{900}{\mega\evperc\squared}$.

After the photon and track selection, a vertex fit is performed to
ensure a common point of origin of all charged tracks.  Next, a
kinematic fit is performed constraining the initial four-momentum of the
$\jpsi$ as well as the mass of the $\piz$ in the $\threepi$ channel.
The mass of the $\eta$ is unconstrained, because the
$M_{\gamma\gamma/\pip\pim\piz}$ spectrum is used to determine the
number of signal events.  If there are multiple candidates per event,
only the candidate with the minimum $\chi^2$ value of the kinematic fit is
selected.  A very loose requirement on the $\chi^2$ value is used to
suppress background events.

\section{\label{sec::BKGDATA}Background studies}

To identify possible background contributions from other $\jpsi$
decays, the inclusive MC sample is used.  The same selection criteria
as for the signal channel are applied to identify the most relevant
background channels surviving the event selection.

In the $\twogg$ decay channel, a wide variety of background
contributions is found, with most channels only contributing a few
events each.  The most prominent background channels involve either an
intermediate charged or neutral $\Delta$ resonance, or a decay of
$\decay{\jpsi}{\proton\antiproton X}$ with $X$ being a light meson
that decays further into a number of photons. About
$\SI{21}{\percent}$ of background events contain misidentified charged
particles. All background categories are distributed smoothly
throughout the $M_{\gamma\gamma}$ spectrum with no peaking behavior in
the signal region.  The amount of background events remaining in the
signal region is about $\SI{4.3}{\percent}$.

In the $\threepi$ decay channel, three major background sources are
identified. The most abundant channel is $\decay{\jpsi}{\Delta X}$
with $X$ being a light baryon. The second dominant background
contribution is the direct production of the final state,
$\decay{\jpsi}{\proton\antiproton\pip\pim\piz}$. Both channels are
distributed smoothly throughout the $M_{\pip\pim\piz}$ spectrum. On
the other hand, the decay
$\decay{\jpsi}{\proton\antiproton\omega\,(\decay{\omega}{\pip\pim\piz})}$
has a sharp peak at the $\omega$ mass, which is well separated from
the signal region. The events of the remaining channels
($\SI{3.7}{\percent}$ of all background events) are distributed
smoothly as well. The amount of background events remaining
in the signal region is with about $\SI{33}{\percent}$,
significantly higher than for the $\twogg$ final state.

Background contributions from the same signal channel but with other
$\eta$ decays are also studied. In the $\twogg$
decay channel, two events from other $\eta$ decays are found, which is
negligible. In the $\threepi$ decay channel, a significant
peaking background contribution of the process
$\decay{\jpsi}{\proton\antiproton\eta\,(\decay{\eta}{\pip\pim\gamma})}$
is found. The inclusive MC sample is used to estimate the rate and
distribution of these events within the signal region. Based on the ratio of
the $\BF\text{s}$, about $\SI{1.51}{\percent}$ of the reconstructed
events are from this process.

An additional source of background events is the process
$e^+e^-\to\gamma^*\to\proton\antiproton\eta$ without a $\jpsi$ as an
intermediate state.  To determine the number of events from this
source, the continuum data sample taken at the CM energy
$\sqrt{s}=\SI{3.080}{\giga\eV}$ is analyzed. The same selection
criteria as for the signal process are applied, with the exception that
the four-momentum of the initial state in the kinematic fit is adjusted.
The number of background events in the signal region is estimated to
be $N^{3080}_{\rm QED}=\num{310+-18}$ in the $\twogg$ channel and
$N^{3080}_{\rm QED}=\num{49+-8}$ in the $\threepi$ channel. Scaling
those numbers to the luminosity of the $\jpsi$ data set yields a
background contribution of $N^{\twogg}_{\rm QED}=\num{5454(317)}$
events in the $\twogg$ channel and $N^{\threepi}_{\rm
QED}=\num{826+-141}$ events in the $\threepi$ channel.  Since it is
expected that the differences in efficiency and cross section between
the two CM energies are much smaller than the statistical uncertainty,
these factors are neglected.

\section{Efficiency Determination} \label{sec:eff} The reconstruction
efficiency describes the probability that a signal event is
reconstructed in the detector and survives the whole selection chain.
It depends heavily on each event's position in the available PHSP,
being drastically lower in regions that contain one or more charged
particles with low momentum, dropping to nearly zero in regions with
$p_{\proton} \leq \SI{200}{\mega\evperc}$.  Moreover, if the
distribution of events deviates from the simple PHSP distribution, a
simulation that accurately reproduces data is required to determine the
correct efficiency. For this analysis, the framework
ComPWA~\cite{ComPWA} is used. The physics model is described by using
the helicity formalism where $N^*$ resonances as intermediate states
are included.  The fit of the model to data is performed using
events from the $\twogg$ channel only, with the additional constraint
on the $\eta$ mass, since this provides an almost background free
sample. The amplitude structure is not expected to differ between the
two analyzed $\eta$ decay channels, so the model is used to generate a
signal MC sample for both channels.

\begin{figure}
 \includegraphics[trim={0 0.0cm 0.0cm 0.0cm}, width=6cm]{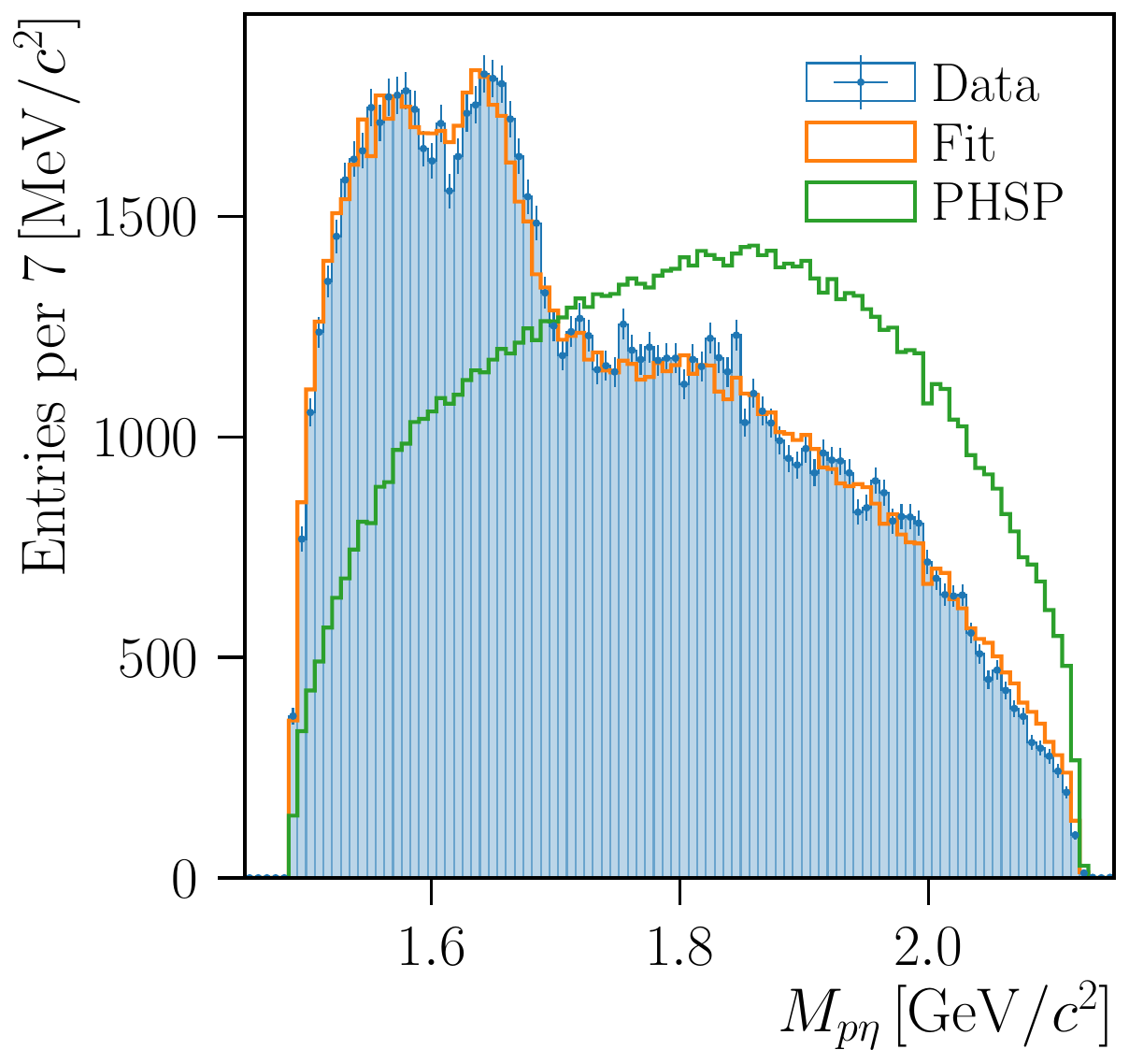}
 \includegraphics[trim={0 0.0cm 0.0cm 0.0cm}, width=6cm]{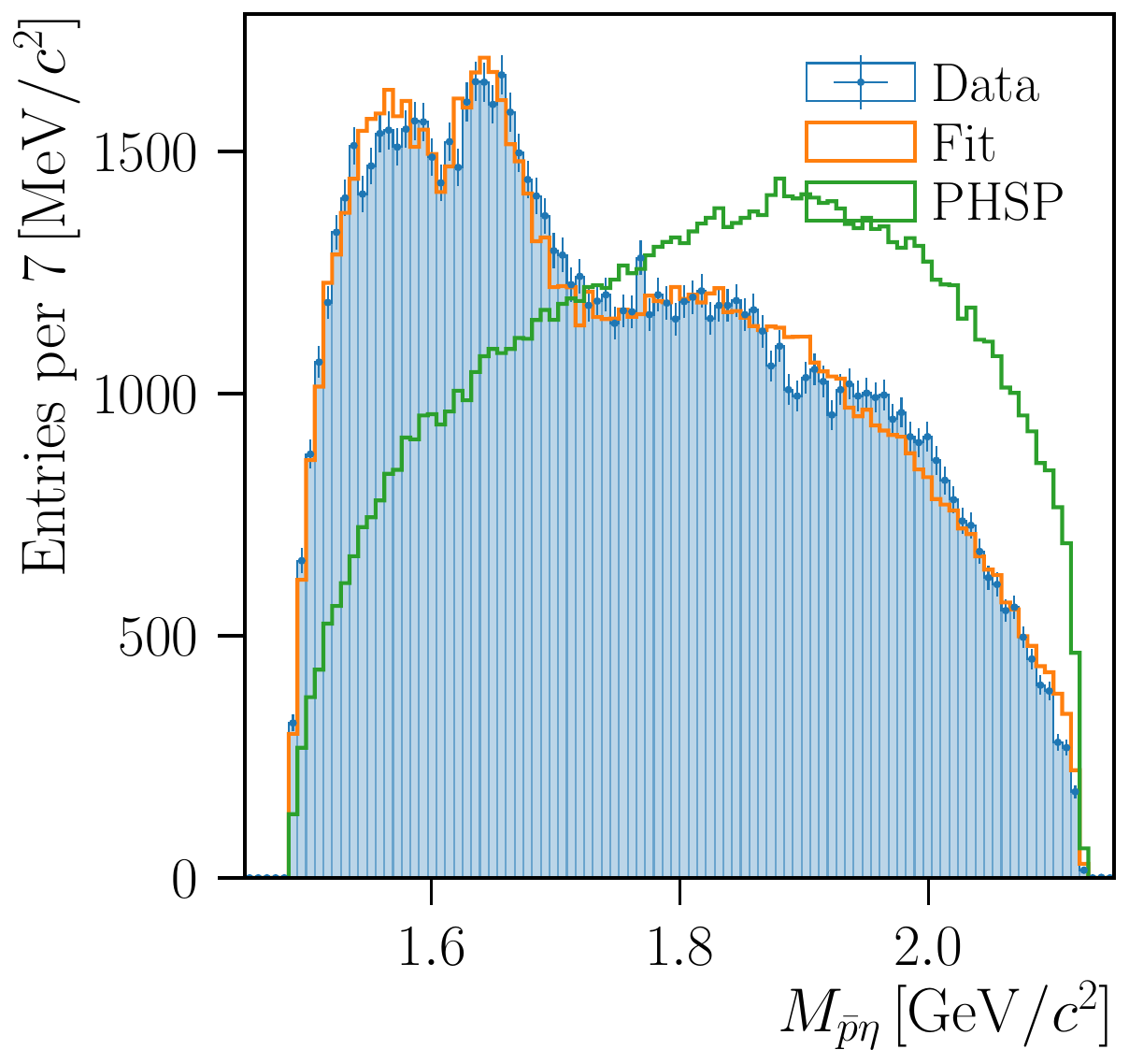}
 \includegraphics[trim={0 0.0cm 0.0cm 0.0cm}, width=6cm]{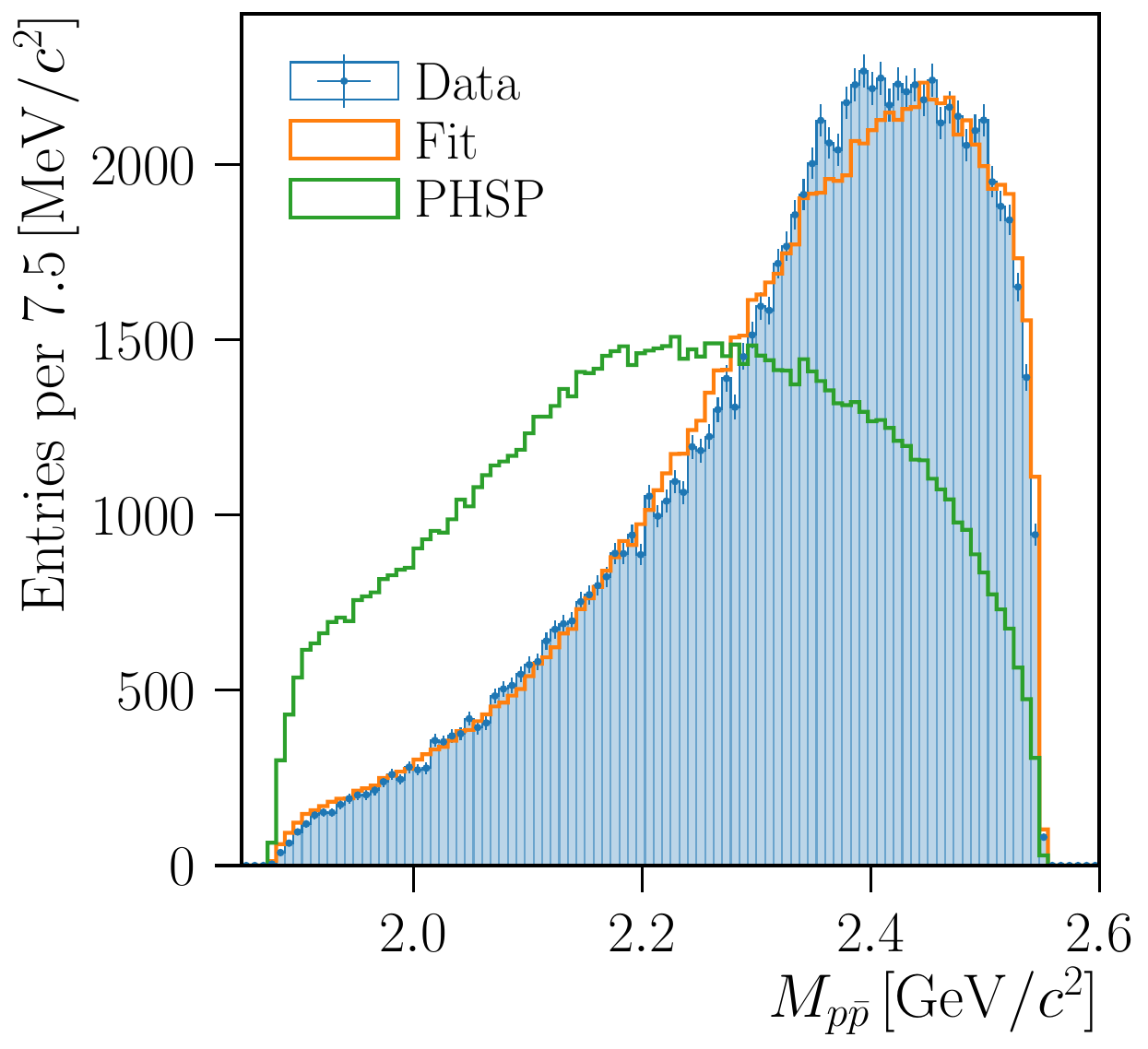}
 \caption{The $M_{\proton\eta}$, $M_{\antiproton\eta}$, and
 $M_{\proton\antiproton}$ distributions in the
 $\twogg$ decay channel
from data (blue dots). The orange histogram shows the
 amplitude model and the green histogram the distribution of the
 PHSP distributed MC sample.}
        \label{fig:Dalitz}
\end{figure}

Figure~\ref{fig:Dalitz} shows the distributions of the invariant mass
of all three sub-systems, $M_{\proton\eta}$, $M_{\antiproton\eta}$, and
$M_{\proton\antiproton}$, together with the amplitude model and the
PHSP distributed MC sample of the three-body final state. For all distributions, the
amplitude model, which includes seven $N^*$ resonances as intermediate
states in the $\antiproton\eta$ and $\proton\eta$ sub-systems, provides
a good description of the data. In particular, the double peak 
structure close to threshold dominating the whole distribution 
could be described well by a destructive 
interference of two $N^*$ resonances, the $N(1535)$ and the $N(1650)$.
The large deviation in the
$\proton\antiproton$ sub-system is described by the reflection caused
by the $N^*$ resonances. The amplitude model describes the density of the
events in the available phase space well, and thus the efficiency
is determined correctly. 

The reconstruction efficiency is calculated with $\epsilon_{\rm
rec}={N_{\rm rec}}/{N_{\rm gen}}$, where $N_{\rm rec}$ represents the number
of reconstructed events and $N_{\rm gen}$ denotes the number of generated
events. The reconstruction efficiency in the $\twogg$
decay channel is determined to be $\epsilon_{\rm rec} =
\SI{44.17(4)}{\percent}$. In the $\threepi$ decay channel,
the efficiency is $\epsilon_{\rm rec} = \SI{16.40(4)}{\percent}$,
which is considerably lower due to the additional charged particles
from the $\eta$ decay, which have comparatively low momentum.

\section{\label{sec::BFCalculation} Branching Fraction}
The branching fraction $\BF$ of the signal decay is calculated by

\begin{equation*} 
    \BF(\ppbareta)=\frac{N_{\rm Sig}}{N_{J\!/\!\psi}}\cdot\frac{1}{\epsilon_{\rm rec}}\cdot\frac{1}{\prod_i{\BF}_i} ,
\end{equation*}

\noindent where $N_{\rm Sig}$ is the number of signal events,
$N_{\jpsi}$ the number of $\jpsi$ events, $\epsilon_{\rm rec}$ the
reconstruction efficiency, and $\prod_i{\BF}_i$ the product of the
branching fractions of the intermediate states, either $\BF(\twogg)$
or $\BF(\threepi) \cdot \BF(\pizero)$.

\begin{figure}
        \includegraphics[trim={1.5cm 1.3cm 0.0 0.0}, width=5cm]{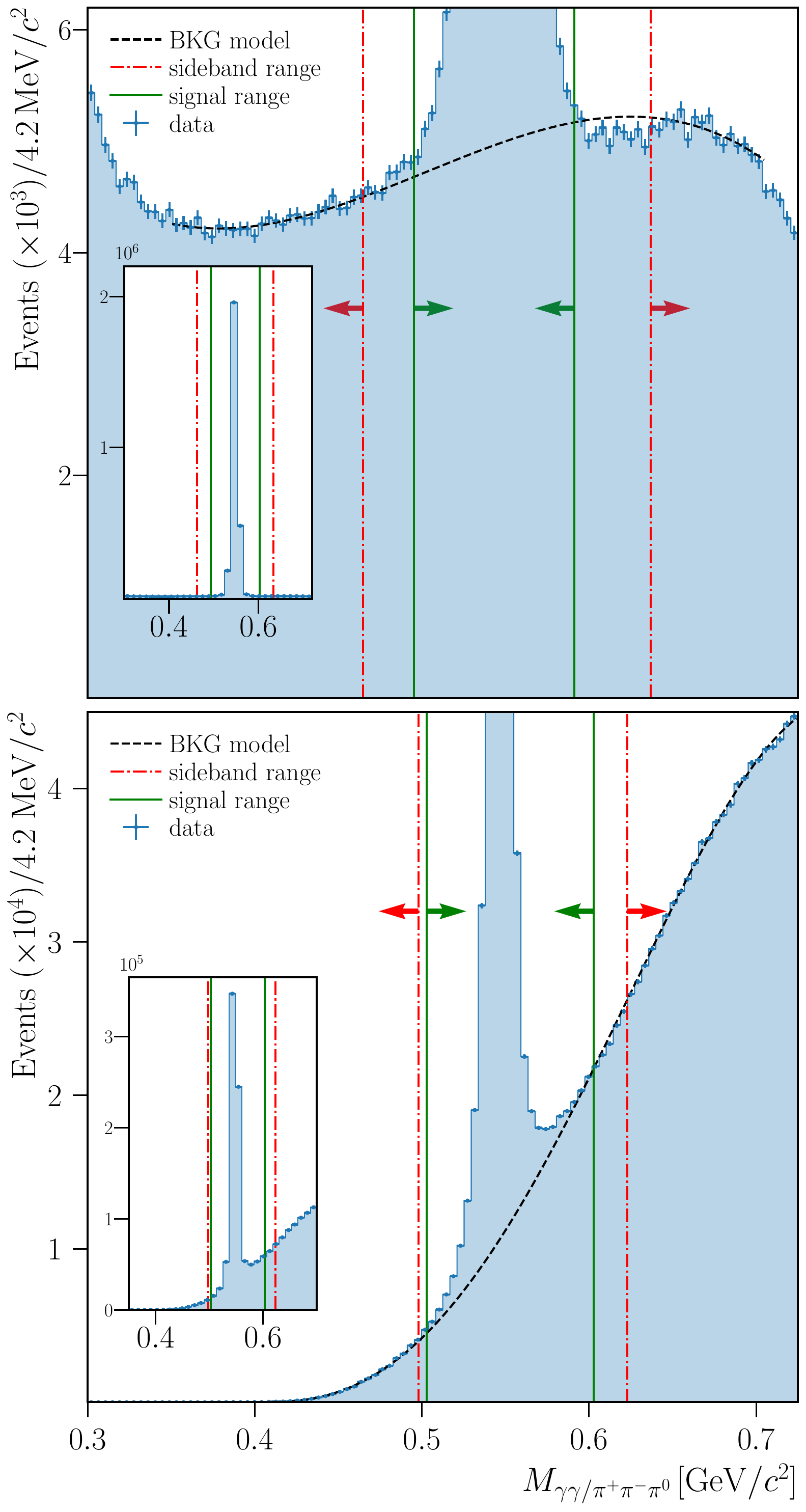}
        \caption{Reconstructed $M_{\gamma\gamma}$ distribution of the $\twogg$ decay channel (top) and reconstructed
        $M_{\pip\pim\piz}$ distribution of the
        $\threepi$ decay channel (bottom).
        The dashed black line describes the background (BKG) model.
        The dashed-dotted red lines mark the boundaries of the sideband region and the solid green lines the boundaries of the signal region.
        The insets show the complete distributions of the same data.}
        \label{fig:Fit}
\end{figure}

The number of signal events $N_{\rm Sig}$ is determined by counting
the number of $\eta$ candidates in the signal region of the
$M_{\gamma\gamma}$ or $M_{\pip\pim\piz}$ distributions, after
subtracting the estimated number of background events (see
Fig.~\ref{fig:Fit}).

In the $\twogg$ decay channel, the signal region is defined as
$\SI{492}{\mega\evperc\squared} \leq M_{\gamma\gamma} \leq
\SI{587}{\mega\evperc\squared}$ (inside the green lines in
Fig.~\ref{fig:Fit}a). The sideband regions are defined as
$\SI{350}{\mega\evperc\squared} \leq M_{\gamma\gamma} \leq
\SI{462}{\mega\evperc\squared}$ and $\SI{632}{\mega\evperc\squared}
\leq M_{\gamma\gamma} \leq \SI{700}{\mega\evperc\squared}$ (outside
the red dashed-dotted lines in Fig.~\ref{fig:Fit}a). To estimate the
number of background events, the sideband regions of the
$M_{\gamma\gamma}$ distribution are fitted with a third-order
Chebychev function to describe the background shape, which is then
interpolated to the signal region to calculate the background yield in
that region. The fit yields \num{11.20(4)e4} background events in the
signal region. Subtracting those as well as the expected \num{5454(317)} QED
background events from the total number of \num{271.60(16)e4}
events in the signal region gives the yield of $\num{\twoggnSig(\twoggnSigErr)e4}$ signal
events.

The fit procedure used in the $\threepi$ decay channel is
similar. The total number of events in the signal region
($\SI{502}{\mega\evperc\squared} \leq M_{\pip\pim\piz} \leq
\SI{602}{\mega\evperc\squared}$, inside the green lines in
Fig.~\ref{fig:Fit}b) is \num{87.68(9)e4}. A fourth-order Chebychev
function is fitted to the background distribution of the sideband
regions ($\SI{407}{\mega\evperc\squared} \leq M_{\pip\pim\piz} \leq
\SI{492}{\mega\evperc\squared}$ and $\SI{622}{\mega\evperc\squared}
\leq M_{\pip\pim\piz} \leq \SI{725}{\mega\evperc\squared}$, outside
the red dashed-dotted lines in Fig.~\ref{fig:Fit}b), which yields
\num{28.37(4)e4} background events in the signal region. After
subtracting the yield of the background polynomial, the \num{826(141)}
QED background events, and the estimated number of \num{0.90(2)e4}
$\decay{\eta}{\pip\pim\gamma}$ events, the signal yield is \num{58.33(10)e4} events.

With the numbers of signal events the branching fractions are \\

\noindent
$ \mathcal{B}(\decay{\jpsi}{\proton\antiproton\eta(\decay{\eta}{\gamma\gamma})}) = (\twoggvalue\pm\num{\twoggstat})\times\,10^{-3}$ ,\\

\noindent
$ \mathcal{B}(\decay{\jpsi}{\proton\antiproton\eta(\decay{\eta}{\pip\pim\piz})}) = (\threepivalue\pm\num{\threepistat})\times\,10^{-3}$ .\\

\noindent
The uncertainty reflects the statistical uncertainty from the number of signal events only.
Table~\ref{tab:calcParameter} shows the complete list of all relevant parameters.

\begin{table}[h!]
    \centering
    \caption{The parameters used for the calculation of the branching fraction $\mathcal{B}$ measurements.}
    \label{tab:calcParameter}
       \begin{tabular}{lrl}
   \\
    \hline\hline
    parameter & \multicolumn{1}{r}{value} 
    \\\hline
    $N_{\jpsi}$  \cite{nJpsi} & $\num{10087(44)e6}$
    \\
    $\BF(\twogg)$ \cite{pdg} &$\SI{39.36+-0.18}{\percent}$
    \\
    $\BF(\threepi)$ \cite{pdg} &$\SI{23.02+-0.25}{\percent}$
    \\
    $\BF(\piz\to\gamma\gamma)$ \cite{pdg} &$\SI{98.823+-0.034}{\percent}$
    \\\hline \hline
    $N_\text{Sig}(\twogg)$  & $\num{\twoggnSig(\twoggnSigErr)e4}$
    \\
    $\epsilon_\text{rec}$ & $\SI{44.17(4)}{\percent}$
    \\\hline \hline
    $N_\text{Sig}(\threepi)$  & $\num{\threepinSig(\threepinSigErr)e4}$
    \\
    $\epsilon_\text{rec}$ & $\SI{16.40(4)}{\percent}$
    \\\hline  \hline
\end{tabular}
\end{table}

\section{Study of threshold enhancement} 
\begin{figure}[b]
        \includegraphics[trim={1.5cm 1.1cm 0.0 0.0}, width=6.0cm]{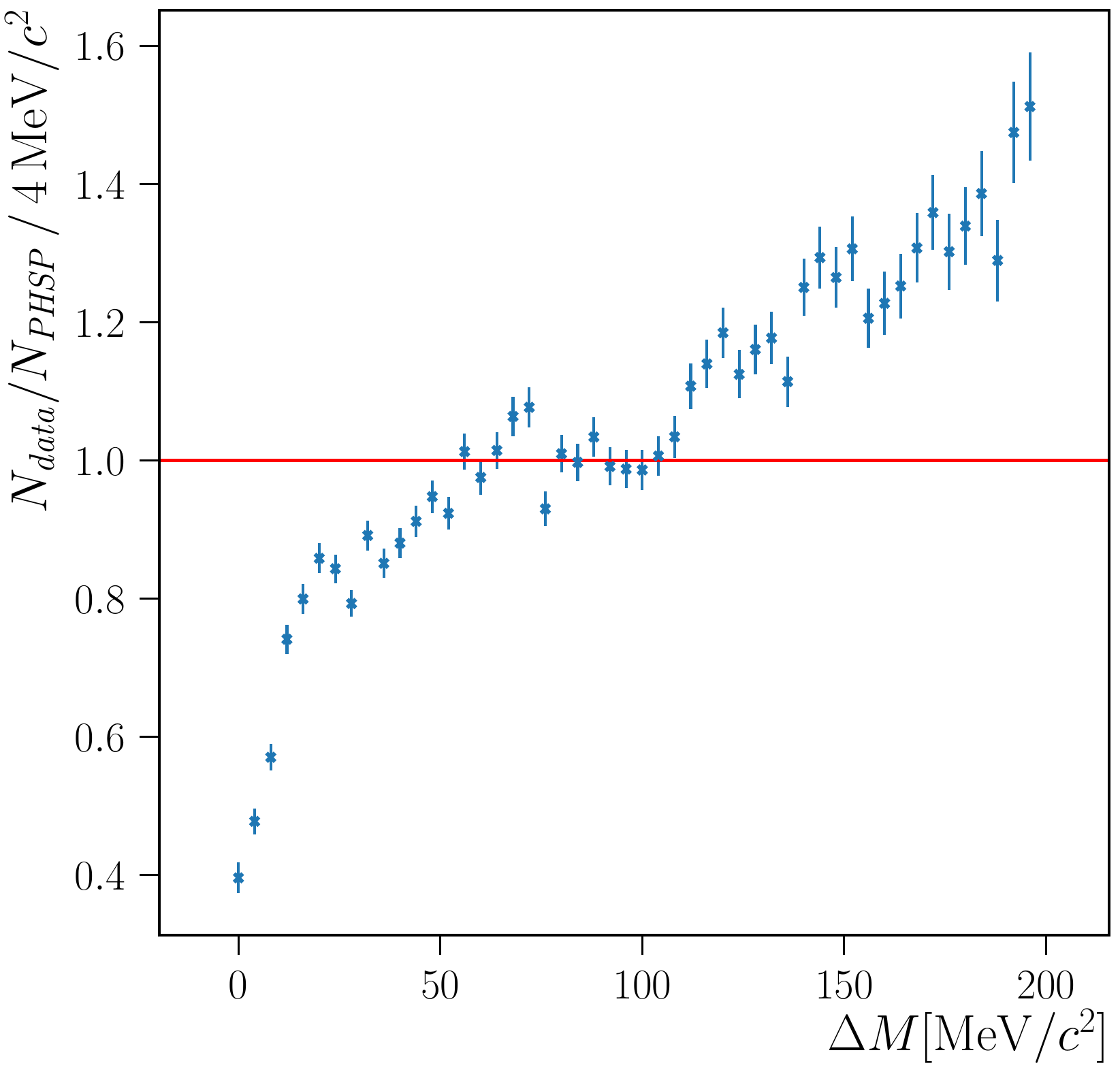}
        \caption{Ratio between efficiency corrected $\ppbareta$ data
        events $N_{\it{data}}$ and with PHSP model generated MC events
        $N_{\rm PHSP}$ versus the mass difference to the
        production threshold ${\Delta}M =
        M_{\proton\antiproton}-2m_{\proton}$. Ratio of 1.0 is
        indicated by the red line.}
        \label{fig:threshold}
\end{figure}

As shown in
Fig.~\ref{fig:Dalitz}, the dynamics in the decay channel is dominated
by processes like $\decay{\jpsi}{\antiproton
N^{*}(\decay{N^{*}}{\proton\eta})}+c.c.$, with strong contributions of
$N^{*}$ resonances with relatively low mass. These contributions would
be considered as background contributions for the study of a possible
threshold enhancement in the $\proton\antiproton$ system. Therefore,
the kinematic regions of $M^2(\proton\eta) \geq
\SI{3.6}{\giga\eV\squared\per\clight\tothe{4}}$ and
$M^2(\antiproton\eta) \geq
\SI{3.6}{\giga\eV\squared\per\clight\tothe{4}}$ are chosen for this
study, because they do not show any obvious resonance
contributions. Additionally, only events that satisfy
$|M_{\gamma\gamma} -m_{\eta}| \leq \SI{20}{\mega\evperc\squared}$ are
considered to reduce background contributions to a level of
\SI{1.8}{\percent}. The impact of these requirements on the one
dimensional distribution of the invariant mass
$M_{\antiproton\proton}$ is substantial. Therefore, the ratio between
the efficiency-corrected data distribution and the generated
distribution of the PHSP MC data set is shown in
Fig.~\ref{fig:threshold} as a function of the mass difference
${\Delta}M$ from the threshold.  The ratio should be equal to 1 if no
contribution from threshold enhancement or $N^*$ resonances is
present. In the low ${\Delta}M =
M_{\proton\antiproton}\,-\,2m_{\proton}$ region, a ratio of greater than 1
would be expected in the presence of a threshold enhancement. In fact, the
opposite behavior is observed, with the ratio between data and MC
simulation being smaller than one in the vicinity of the threshold. 
This suggests either the absence of a threshold enhancement
in the $\proton\antiproton$ system, consistent with the
previous results \cite{ppbareta2001}, or a complex interplay of the
$N^*$ and $\proton\antiproton$ amplitudes. 

\section{\label{sec::SYSTEMATICS} Systematic uncertainty estimation}

In the following the different sources of the systematic uncertainties are described. 

The systematic uncertainty of the track reconstruction efficiency is
determined using a weighting method, which takes into account the
dependence on the transverse momentum and the $\cos\theta$ of the
tracks, when estimating the difference between data and MC
simulation. The weights are obtained by studying the decay $\jpsi\to
\pi^+\pi^- \proton \antiproton$, which closely resembles the signal
decay.  The weighting is performed individually for every charged
particle type, resulting in a total systematic uncertainty of
\SI{0.49}{\percent} in the $\twogg$ decay channel. In the $\threepi$
decay channel, the uncertainty for the protons is \SI{0.47}{\percent},
and the uncertainty for the pions is \SI{0.78}{\percent}.

The difference in the reconstruction efficiency of photons between
data and MC simulation is studied with the decay channel
$\jpsi\to\gamma\mu^+\mu^-$.  The resulting systematic uncertainty is
$\SI{0.5}{\percent}$ per photon, or a total uncertainty of
$\SI{1}{\percent}$.

The systematic uncertainty of the efficiency related to the particle
identification in the channel with the $\threepi$ decay is determined
also by using the weighting method and the channel $\jpsi\to
\pi^+\pi^- \proton \antiproton$. In this case the dependence on the
momentum and the $\cos\theta$ of the tracks is taken into account
in estimating the difference between data and MC simulation. The
weighting is performed individually for every charged particle type,
resulting in a total systematic uncertainty of \SI{1.02}{\percent}. In
the $\twogg$ decay channel, no particle identification is
used, and therefore no uncertainty is assigned.

The systematic uncertainty introduced by the veto on photon candidates, that are
detected within a $\SI{20}{\degree}$ cone around a charged track ($\Delta\alpha$), is
estimated by varying the requirement by $\pm\SI{3}{\degree}$, which
corresponds to taking into account the measurement of one less or one
additional calorimeter crystal at low $\cos\theta$. The uncertainty is
estimated to be $\SI{0.08}{\percent}$ and $\SI{0.07}{\percent}$ for
the two channels, respectively.

The systematic uncertainty introduced by the requirement on the
$\chi^2$ value of the kinematic fit is estimated varying the
requirement by $\pm\SI{10}{\percent}$ and assigning the largest
difference in $\BF$ as the systematic uncertainty. It is estimated to be
$\SI{0.11}{\percent}$ and $\SI{0.17}{\percent}$ for the two channels,
respectively.

The statistical uncertainty of the reconstruction efficiency is
treated as the systematic uncertainty for the branching fraction, with
$\SI{0.09}{\percent}$ in the $\twogg$ channel and
$\SI{0.25}{\percent}$ in the $\threepi$ channel.

To estimate the systematic effects introduced by the choice of the
boundaries of the signal and sideband ranges, each boundary is
individually varied within $\pm\SI{10}{\percent}$. The exception to
this is the lower boundary of the sideband range in the channel with
the $\threepi$ decay, since it is already placed at the edge of the
available phase space, so it is only varied upwards. The largest
difference in $\BF$ for each category is assigned as the systematic
uncertainty. All uncertainties are checked to see if they are already
covered by statistical fluctuations by using the Barlow
test \cite{barlow}. Using this method, the following uncertainties are
assigned of $\SI{0.01}{\percent}$ $(\SI{0.02}{\percent})$ for the lower
bound of the signal range, $ \SI{0.01}{\percent}$ $(\SI{0.04}{\percent})$
for the upper bound of the signal range, $
\SI{0.03}{\percent}$ $(\SI{0.17}{\percent})$ for the width of the
background window, $\SI{0.08}{\percent}$ $(\SI{0.54}{\percent})$ for the
lower bound of the whole fit range, and
$\SI{0.11}{\percent}$ $(\SI{0.37}{\percent})$ for the upper bound of the
whole fit range, in the $\twogg$ $(\threepi)$ channel.

The shape of the background model is varied by changing the order of
the polynomial function describing the background shape by plus/minus
one order. The largest difference to the nominal value is taken as the
uncertainty, which is $\SI{0.92}{\percent}$ and $\SI{0.38}{\percent}$
for the $\twogg$ and the $\threepi$ channels, respectively.

The systematic uncertainty for the continuum background is calculated
by Gaussian error propagation using the uncertainty of $N_{3080}^{\rm
QED}$ which contributes an uncertainty of $\SI{0.01}{\percent}$ in
both channels.

The systematic uncertainty introduced by the determination of the
amplitude model is estimated by varying the parameters of the model
within the range taken from the covariance matrix. For 1000 different 
sets of parameters the efficiency is determined 
which results in a distribution of efficiency values.
The standard
deviation of this distribution is taken as the systematic
uncertainty which is $\SI{0.05}{\percent}$ for the $\twogg$ channel
and $\SI{0.06}{\percent}$ for the $\threepi$ channel.

For the external parameters such as the total number of $\jpsi$ events
and the branching fractions of the intermediate particles Gaussian
error propagation is used.  For $N_{\jpsi}$ this results in a
systematic uncertainty of \num{0.43}\% \cite{nJpsi}, for
$\mathcal{B}(\twogg)$ in \num{0.46}\%, for $\mathcal{B}(\threepi)$ in
\num{1.09}\%, and for $\mathcal{B}(\piz\to\gamma\gamma)$ in
\num{0.03}\% \cite{pdg}.

The total systematic uncertainties, which are listed in
Table~\ref{tab:sys}, are calculated by summing all
individual uncertainties in quadrature.  The resulting relative systematic uncertainty is
$\SI{1.60}{\percent}$ for the $\twogg$ decay channel and
$\SI{2.24}{\percent}$ for the $\threepi$ decay channel, which
results in the absolute systematic uncertainties of
$\num{\twoggsys}\times\,10^{-3}$ and
$\num{\threepisys}\times\,10^{-3}$,
respectively. Separating the correlated and uncorrelated systematic
uncertainties, $\num{0.018}\times\,10^{-3}$ corresponds in both cases
to the correlated uncertainties.

The two measurements are combined taking into account the correlated
and uncorrelated contributions to the systematic uncertainties of both
channels \cite{combination_formula}. The combined $\BF$ is \\

\noindent
$\BF(\ppbareta)=(\BFcombi\,\pm\,\num{\SigmacombiStat}\,\pm\,\num{\Sigmacombi})\times\,10^{-3}. $
\\

\noindent
The first uncertainty is the combined statistical uncertainty and the second the combined systematic uncertainty of both analyses. 

\begin{table}[t]
  \caption{Systematic uncertainties by source and the total systematic uncertainties. Uncertainties marked with (*) are considered correlated between the two channels.}
  \label{tab:sys}
  \centering
  \begin{tabular}{lcc}
  \\
    \hline\hline
    Source & $\twogg$ & $\threepi$\\
    \hline 
    $\proton\antiproton$ tracks (*) & 0.49 & 0.47 \\ 
    $\pip\pim$ tracks  & - & 0.78 \\ 
    Photons (*)  & 1.00 & 1.00 \\ 
    PID & - & 1.02 \\
    \hline
    $\Delta\alpha$ (*) & 0.08 & 0.07 \\
    Kinematic fit     & \num{0.11} & \num{0.17}\\ 
    Efficiency        & 0.09 & 0.25 \\
    Signal range min & \num{0.01} & \num{0.02}\\
    Signal range max & \num{0.01} & \num{0.04}\\
    Background window & \num{0.03} & \num{0.17}\\
    Fit range min & \num{0.08} & \num{0.54}\\
    Fit range max & \num{0.11} & \num{0.37}\\
    Background model & 0.92 & 0.38 \\
    QED background & 0.01 & 0.01 \\
    Amplitude model & 0.05 & 0.06 \\
    \hline
    $N_{\jpsi}$ (*) & 0.43 & 0.43 \\
    $\BF_{\eta}$ & 0.46 & 1.09 \\ 
    $\BF_{\piz}$ & - & 0.03  \\ 
    \hline 
    Total    & 1.60 & 2.24\\
    \hline\hline
  \end{tabular}
\end{table}

\section{\label{sec::SUMMARY}Summary}

This paper describes the most precise measurement to date of the branching
fraction of the decay $\ppbareta$, using the BESIII data set of
\num{10087(44)e6} $\jpsi$ events. Two different $\eta$ final states,
$\twogg$ and $\threepi$, are used for this analysis. The single
branching fractions are determined to be \\

\noindent
$\mathcal{B}(\decay{\jpsi}{\proton\antiproton\eta(\decay{\eta}{\gamma\gamma})}) = $ \\
\hspace*{24mm} $(\twoggvalue\,\pm\,\num{\twoggstat}\,\pm\,\num{0.018}\,\pm\,\num{0.016})\times\,10^{-3}$, \\

\noindent
$\mathcal{B}(\decay{\jpsi}{\proton\antiproton\eta(\decay{\eta}{\pip\pim\piz})}) = $ \\
\hspace*{24mm} $(\threepivalue\,\pm\,\num{\threepistat}\,\pm\,\num{0.018}\,\pm\,\num{0.029})\times\,10^{-3}$, \\

\noindent where the first uncertainty is statistical and the second and third corresponds to the correlated and uncorrelated systematic uncertainties, respectively. The difference between the two measurements 
is about $2.0 \sigma$  taking into account all uncorrelated
uncertainties. Therefore, the measurements agree within their
uncertainties. A small difference between the branching fractions of
these two decays was already observed before by BESII, but the other
way around \cite{oldbes}.

The combined branching fraction is \\

\noindent
$\BF(\ppbareta)=(\BFcombi\,\pm\,\num{\SigmacombiStat}\,\pm\,\num{\Sigmacombi})\times\,10^{-3}, $ \\

\noindent where the first uncertainty is the combined statistical
uncertainty and the second one the combined systematic uncertainty of
both analyses. Correlations between both are taken into account. The
combined result differs from the previous world average by $4.1
\sigma$. Former experiments used a pure three-body PHSP model for
the determination of the global reconstruction efficiency. For this
analysis an amplitude analysis is performed to obtain better data/MC
consistency. This causes part of the observed difference with the old
experiments. 
The largest deviation from the pure three-body PHSP distribution was
found to be caused by the destructive interference of the $N(1535)$
and the $N(1650)$ resonances.

In addition, the $\proton\antiproton$ threshold region is studied. No
evidence for any threshold enhancement in this channel is observed. 

\section*{Acknowledgments}

The BESIII Collaboration thanks the staff of BEPCII and the IHEP computing center for their strong support. This work is 
supported in part by National Key R\&D Program of China under Contracts Nos. 2020YFA0406300, 
2020YFA0406400; National Natural Science Foundation of China (NSFC) under Contracts Nos. 11635010, 11735014, 11835012, 11935015, 11935016, 11935018, 11961141012, 12025502, 12035009, 12035013, 12061131003, 12192260, 12192261, 12192262, 12192263, 12192264, 12192265, 12221005, 12225509, 12235017; the Chinese Academy of Sciences (CAS) Large-Scale Scientific Facility Program; the CAS Center for Excellence in Particle Physics (CCEPP); Joint Large-Scale Scientific Facility Funds of the NSFC and CAS under Contract No. U1832207; CAS Key Research Program of Frontier Sciences under Contracts Nos. QYZDJ-SSW-SLH003, QYZDJ-SSW-SLH040; 100 Talents Program of CAS; The Institute of Nuclear and Particle Physics (INPAC) and Shanghai Key Laboratory for Particle Physics and Cosmology; European Union's Horizon 2020 research and innovation programme under Marie Sklodowska-Curie grant agreement 
under Contract No. 894790; German Research Foundation DFG under Contracts Nos. 455635585, Collaborative Research 
Center CRC 1044, FOR5327, GRK 2149; Istituto Nazionale di Fisica Nucleare, Italy; Ministry of Development of Turkey under Contract No. DPT2006K-120470; National Research Foundation of Korea under Contract No. NRF-2022R1A2C1092335; National Science and Technology fund of Mongolia; National Science Research and Innovation Fund (NSRF) via the Program Management Unit for Human Resources \& Institutional Development, Research and Innovation of Thailand under Contract No. 
B16F640076; Polish National Science Centre under Contract No. 2019/35/O/ST2/02907; 
The Swedish Research Council; U. S. Department of Energy under Contract No. DE-FG02-05ER41374

\FloatBarrier

\bibliography{Bibliography}

\providecommand{\noopsort}[1]{}\providecommand{\singleletter}[1]{#1}%
\begin{thebibliography}{33}%
\makeatletter
\providecommand \@ifxundefined [1]{%
 \@ifx{#1\undefined}
}%
\providecommand \@ifnum [1]{%
 \ifnum #1\expandafter \@firstoftwo
 \else \expandafter \@secondoftwo
 \fi
}%
\providecommand \@ifx [1]{%
 \ifx #1\expandafter \@firstoftwo
 \else \expandafter \@secondoftwo
 \fi
}%
\providecommand \natexlab [1]{#1}%
\providecommand \enquote  [1]{``#1''}%
\providecommand \bibnamefont  [1]{#1}%
\providecommand \bibfnamefont [1]{#1}%
\providecommand \citenamefont [1]{#1}%
\providecommand \href@noop [0]{\@secondoftwo}%
\providecommand \href [0]{\begingroup \@sanitize@url \@href}%
\providecommand \@href[1]{\@@startlink{#1}\@@href}%
\providecommand \@@href[1]{\endgroup#1\@@endlink}%
\providecommand \@sanitize@url [0]{\catcode `\\12\catcode `\$12\catcode
  `\&12\catcode `\#12\catcode `\^12\catcode `\_12\catcode `\%12\relax}%
\providecommand \@@startlink[1]{}%
\providecommand \@@endlink[0]{}%
\providecommand \url  [0]{\begingroup\@sanitize@url \@url }%
\providecommand \@url [1]{\endgroup\@href {#1}{\urlprefix }}%
\providecommand \urlprefix  [0]{URL }%
\providecommand \Eprint [0]{\href }%
\providecommand \doibase [0]{https://doi.org/}%
\providecommand \selectlanguage [0]{\@gobble}%
\providecommand \bibinfo  [0]{\@secondoftwo}%
\providecommand \bibfield  [0]{\@secondoftwo}%
\providecommand \translation [1]{[#1]}%
\providecommand \BibitemOpen [0]{}%
\providecommand \bibitemStop [0]{}%
\providecommand \bibitemNoStop [0]{.\EOS\space}%
\providecommand \EOS [0]{\spacefactor3000\relax}%
\providecommand \BibitemShut  [1]{\csname bibitem#1\endcsname}%
\let\auto@bib@innerbib\@empty
\bibitem [{\citenamefont {Ronniger}\ and\ \citenamefont
  {Metsch}(2011)}]{Ronniger:2011td}%
  \BibitemOpen
  \bibfield  {author} {\bibinfo {author} {\bibfnamefont {M.}~\bibnamefont
  {Ronniger}}\ and\ \bibinfo {author} {\bibfnamefont {B.~C.}\ \bibnamefont
  {Metsch}},\ }\href {https://doi.org/10.1140/epja/i2011-11162-8} {\bibfield
  {journal} {\bibinfo  {journal} {Eur. Phys. J. A}\ }\textbf {\bibinfo {volume}
  {47}},\ \bibinfo {pages} {162} (\bibinfo {year} {2011})}\BibitemShut
  {NoStop}%
\bibitem [{\citenamefont {Edwards}\ \emph {et~al.}(2011)\citenamefont {Edwards}
  \emph {et~al.}}]{Edwards:2011jj}%
  \BibitemOpen
  \bibfield  {author} {\bibinfo {author} {\bibfnamefont {R.~G.}\ \bibnamefont
  {Edwards}} \emph {et~al.},\ }\href
  {https://doi.org/10.1103/PhysRevD.84.074508} {\bibfield  {journal} {\bibinfo
  {journal} {Phys. Rev. D}\ }\textbf {\bibinfo {volume} {84}},\ \bibinfo
  {pages} {074508} (\bibinfo {year} {2011})}\BibitemShut {NoStop}%
\bibitem [{\citenamefont {Workman}\ \emph {et~al.}(2022)\citenamefont {Workman}
  \emph {et~al.}}]{pdg}%
  \BibitemOpen
  \bibfield  {author} {\bibinfo {author} {\bibfnamefont {R.~L.}\ \bibnamefont
  {Workman}} \emph {et~al.} (\bibinfo {collaboration} {Particle Data Group}),\
  }\href@noop {} {\bibfield  {journal} {\bibinfo  {journal} {Prog. Theor. Exp.
  Phys.}\ ,\ \bibinfo {pages} {083C01}} (\bibinfo {year} {2022})}\BibitemShut
  {NoStop}%
\bibitem [{\citenamefont {Bai}\ \emph {et~al.}(2003)\citenamefont {Bai} \emph
  {et~al.}}]{ppbar2003}%
  \BibitemOpen
  \bibfield  {author} {\bibinfo {author} {\bibfnamefont {J.~Z.}\ \bibnamefont
  {Bai}} \emph {et~al.} (\bibinfo {collaboration} {BES Collaboration}),\
  }\href@noop {} {\bibfield  {journal} {\bibinfo  {journal} {Phys. Rev. Lett.}\
  }\textbf {\bibinfo {volume} {91}},\ \bibinfo {pages} {022001} (\bibinfo
  {year} {2003})}\BibitemShut {NoStop}%
\bibitem [{\citenamefont {Ablikim}\ \emph {et~al.}(2012)\citenamefont {Ablikim}
  \emph {et~al.}}]{ppbar2012}%
  \BibitemOpen
  \bibfield  {author} {\bibinfo {author} {\bibfnamefont {M.}~\bibnamefont
  {Ablikim}} \emph {et~al.} (\bibinfo {collaboration} {BESIII Collaboration}),\
  }\href@noop {} {\bibfield  {journal} {\bibinfo  {journal} {Phys. Rev. Lett.}\
  }\textbf {\bibinfo {volume} {108}},\ \bibinfo {pages} {182001} (\bibinfo
  {year} {2012})}\BibitemShut {NoStop}%
\bibitem [{\citenamefont {Bai}\ \emph {et~al.}(2001)\citenamefont {Bai} \emph
  {et~al.}}]{ppbareta2001}%
  \BibitemOpen
  \bibfield  {author} {\bibinfo {author} {\bibfnamefont {J.}~\bibnamefont
  {Bai}} \emph {et~al.} (\bibinfo {collaboration} {BES Collaboration}),\ }\href
  {https://doi.org/10.1016/s0370-2693(01)00605-0} {\bibfield  {journal}
  {\bibinfo  {journal} {Phys. Lett. B}\ }\textbf {\bibinfo {volume} {510}},\
  \bibinfo {pages} {75} (\bibinfo {year} {2001})}\BibitemShut {NoStop}%
\bibitem [{\citenamefont {Ablikim}\ \emph {et~al.}(2008)\citenamefont {Ablikim}
  \emph {et~al.}}]{ppbaromega2008}%
  \BibitemOpen
  \bibfield  {author} {\bibinfo {author} {\bibfnamefont {M.}~\bibnamefont
  {Ablikim}} \emph {et~al.} (\bibinfo {collaboration} {BESIII Collaboration}),\
  }\href {https://doi.org/10.1140/epjc/s10052-007-0467-4} {\bibfield  {journal}
  {\bibinfo  {journal} {Eur. Phys. J. C}\ }\textbf {\bibinfo {volume} {53}},\
  \bibinfo {pages} {15} (\bibinfo {year} {2008})}\BibitemShut {NoStop}%
\bibitem [{\citenamefont {Ablikim}\ \emph {et~al.}(2013)\citenamefont {Ablikim}
  \emph {et~al.}}]{ppbaromega2013}%
  \BibitemOpen
  \bibfield  {author} {\bibinfo {author} {\bibfnamefont {M.}~\bibnamefont
  {Ablikim}} \emph {et~al.} (\bibinfo {collaboration} {BESIII Collaboration}),\
  }\href {https://doi.org/10.1103/PhysRevD.87.112004} {\bibfield  {journal}
  {\bibinfo  {journal} {Phys. Rev. D}\ }\textbf {\bibinfo {volume} {87}},\
  \bibinfo {pages} {112004} (\bibinfo {year} {2013})}\BibitemShut {NoStop}%
\bibitem [{\citenamefont {Alexander}\ \emph {et~al.}(2010)\citenamefont
  {Alexander} \emph {et~al.}}]{psiToppbarXCleo}%
  \BibitemOpen
  \bibfield  {author} {\bibinfo {author} {\bibfnamefont {J.~P.}\ \bibnamefont
  {Alexander}} \emph {et~al.} (\bibinfo {collaboration} {CLEO Collaboration}),\
  }\href {https://doi.org/10.1103/PhysRevD.82.092002} {\bibfield  {journal}
  {\bibinfo  {journal} {Phys. Rev. D}\ }\textbf {\bibinfo {volume} {82}},\
  \bibinfo {pages} {092002} (\bibinfo {year} {2010})}\BibitemShut {NoStop}%
\bibitem [{\citenamefont {Ablikim}\ \emph {et~al.}(2016)\citenamefont {Ablikim}
  \emph {et~al.}}]{gammaetaprpipi}%
  \BibitemOpen
  \bibfield  {author} {\bibinfo {author} {\bibfnamefont {M.}~\bibnamefont
  {Ablikim}} \emph {et~al.} (\bibinfo {collaboration} {BESIII Collaboration}),\
  }\href@noop {} {\bibfield  {journal} {\bibinfo  {journal} {Phys. Rev. Lett.}\
  }\textbf {\bibinfo {volume} {117}},\ \bibinfo {pages} {042002} (\bibinfo
  {year} {2016})}\BibitemShut {NoStop}%
\bibitem [{\citenamefont {Ablikim}\ \emph {et~al.}(2011)\citenamefont {Ablikim}
  \emph {et~al.}}]{X1835-2011}%
  \BibitemOpen
  \bibfield  {author} {\bibinfo {author} {\bibfnamefont {M.}~\bibnamefont
  {Ablikim}} \emph {et~al.} (\bibinfo {collaboration} {BESIII Collaboration}),\
  }\href {https://doi.org/10.1103/PhysRevLett.106.072002} {\bibfield  {journal}
  {\bibinfo  {journal} {Phys. Rev. Lett.}\ }\textbf {\bibinfo {volume} {106}},\
  \bibinfo {pages} {072002} (\bibinfo {year} {2011})}\BibitemShut {NoStop}%
\bibitem [{\citenamefont {Ablikim}\ \emph {et~al.}(2015)\citenamefont {Ablikim}
  \emph {et~al.}}]{gammaKSKSeta2015}%
  \BibitemOpen
  \bibfield  {author} {\bibinfo {author} {\bibfnamefont {M.}~\bibnamefont
  {Ablikim}} \emph {et~al.} (\bibinfo {collaboration} {BESIII Collaboration}),\
  }\href@noop {} {\bibfield  {journal} {\bibinfo  {journal} {Phys. Rev. Lett.}\
  }\textbf {\bibinfo {volume} {115}},\ \bibinfo {pages} {091803} (\bibinfo
  {year} {2015})}\BibitemShut {NoStop}%
\bibitem [{\citenamefont {Kang}\ \emph {et~al.}(2015)\citenamefont {Kang},
  \citenamefont {Haidenbauer},\ and\ \citenamefont {Meissner}}]{ppbar_bound1}%
  \BibitemOpen
  \bibfield  {author} {\bibinfo {author} {\bibfnamefont {X.-W.}\ \bibnamefont
  {Kang}}, \bibinfo {author} {\bibfnamefont {J.}~\bibnamefont {Haidenbauer}},\
  and\ \bibinfo {author} {\bibfnamefont {U.-G.}\ \bibnamefont {Meissner}},\
  }\href@noop {} {\bibfield  {journal} {\bibinfo  {journal} {Phys. Rev. D}\
  }\textbf {\bibinfo {volume} {91}},\ \bibinfo {pages} {074003} (\bibinfo
  {year} {2015})}\BibitemShut {NoStop}%
\bibitem [{\citenamefont {Dedonder}\ \emph {et~al.}(2018)\citenamefont
  {Dedonder}, \citenamefont {Loiseau},\ and\ \citenamefont
  {Wycech}}]{ppbar_bound2}%
  \BibitemOpen
  \bibfield  {author} {\bibinfo {author} {\bibfnamefont {J.-P.}\ \bibnamefont
  {Dedonder}}, \bibinfo {author} {\bibfnamefont {B.}~\bibnamefont {Loiseau}},\
  and\ \bibinfo {author} {\bibfnamefont {S.}~\bibnamefont {Wycech}},\ }\href
  {https://doi.org/10.1103/PhysRevC.97.065206} {\bibfield  {journal} {\bibinfo
  {journal} {Phys. Rev. C}\ }\textbf {\bibinfo {volume} {97}},\ \bibinfo
  {pages} {065206} (\bibinfo {year} {2018})}\BibitemShut {NoStop}%
\bibitem [{\citenamefont {Li}(2006)}]{ppbar_glue1}%
  \BibitemOpen
  \bibfield  {author} {\bibinfo {author} {\bibfnamefont {B.~A.}\ \bibnamefont
  {Li}},\ }\href {https://doi.org/10.1103/PhysRevD.74.034019} {\bibfield
  {journal} {\bibinfo  {journal} {Phys. Rev. D}\ }\textbf {\bibinfo {volume}
  {74}},\ \bibinfo {pages} {034019} (\bibinfo {year} {2006})}\BibitemShut
  {NoStop}%
\bibitem [{\citenamefont {Kochelev}\ and\ \citenamefont
  {Min}(2006)}]{ppbar_glue2}%
  \BibitemOpen
  \bibfield  {author} {\bibinfo {author} {\bibfnamefont {N.}~\bibnamefont
  {Kochelev}}\ and\ \bibinfo {author} {\bibfnamefont {D.-P.}\ \bibnamefont
  {Min}},\ }\href
  {https://doi.org/https://doi.org/10.1016/j.physletb.2005.11.079} {\bibfield
  {journal} {\bibinfo  {journal} {Phys. Lett. B}\ }\textbf {\bibinfo {volume}
  {633}},\ \bibinfo {pages} {283} (\bibinfo {year} {2006})}\BibitemShut
  {NoStop}%
\bibitem [{\citenamefont {Bloms}(2020)}]{threshold_overview}%
  \BibitemOpen
  \bibfield  {author} {\bibinfo {author} {\bibfnamefont {J.}~\bibnamefont
  {Bloms}},\ }\href {https://doi.org/10.1063/5.0008592} {\bibfield  {journal}
  {\bibinfo  {journal} {AIP Conference Proceedings}\ }\textbf {\bibinfo
  {volume} {2249}},\ \bibinfo {pages} {030029} (\bibinfo {year}
  {2020})}\BibitemShut {NoStop}%
\bibitem [{\citenamefont {Dmitriev}\ \emph {et~al.}(2016)\citenamefont
  {Dmitriev} \emph {et~al.}}]{ppbar_finalState}%
  \BibitemOpen
  \bibfield  {author} {\bibinfo {author} {\bibfnamefont {V.}~\bibnamefont
  {Dmitriev}} \emph {et~al.},\ }\href
  {https://doi.org/10.1016/j.physletb.2016.06.056} {\bibfield  {journal}
  {\bibinfo  {journal} {Phys. Lett. B}\ }\textbf {\bibinfo {volume} {760}},\
  \bibinfo {pages} {139} (\bibinfo {year} {2016})}\BibitemShut {NoStop}%
\bibitem [{\citenamefont {Ablikim}\ \emph {et~al.}(2009)\citenamefont {Ablikim}
  \emph {et~al.}}]{oldbes}%
  \BibitemOpen
  \bibfield  {author} {\bibinfo {author} {\bibfnamefont {M.}~\bibnamefont
  {Ablikim}} \emph {et~al.} (\bibinfo {collaboration} {BESIII Collaboration}),\
  }\href@noop {} {\bibfield  {journal} {\bibinfo  {journal} {Phys. Lett. B}\
  }\textbf {\bibinfo {volume} {676}},\ \bibinfo {pages} {25} (\bibinfo {year}
  {2009})}\BibitemShut {NoStop}%
\bibitem [{\citenamefont {Ablikim}\ \emph {et~al.}(2010)\citenamefont {Ablikim}
  \emph {et~al.}}]{BESIIISYS:2009}%
  \BibitemOpen
  \bibfield  {author} {\bibinfo {author} {\bibfnamefont {M.}~\bibnamefont
  {Ablikim}} \emph {et~al.} (\bibinfo {collaboration} {BESIII Collaboration}),\
  }\href {https://doi.org/https://doi.org/10.1016/j.nima.2009.12.050}
  {\bibfield  {journal} {\bibinfo  {journal} {Nucl. Instr. Meth. A}\ }\textbf
  {\bibinfo {volume} {614}},\ \bibinfo {pages} {345 } (\bibinfo {year}
  {2010})}\BibitemShut {NoStop}%
\bibitem [{\citenamefont {Yu}\ \emph {et~al.}()\citenamefont {Yu} \emph
  {et~al.}}]{Yu:IPAC2016-TUYA01}%
  \BibitemOpen
  \bibfield  {author} {\bibinfo {author} {\bibfnamefont {C.~H.}\ \bibnamefont
  {Yu}} \emph {et~al.},\ }\href@noop {} {\bibinfo  {journal} {BEPCII
  Performance and Beam Dynamics Studies on Luminosity, IPAC2016 (Busan, Korea,
  2016)}\ }\BibitemShut {NoStop}%
\bibitem [{\citenamefont {Guo}\ \emph {et~al.}(2019)\citenamefont {Guo} \emph
  {et~al.}}]{etof::Guo2019}%
  \BibitemOpen
\bibfield  {journal} {  }\bibfield  {author} {\bibinfo {author} {\bibfnamefont
  {X.~L.}\ \bibnamefont {Guo}} \emph {et~al.},\ }\href@noop {} {\bibfield
  {journal} {\bibinfo  {journal} {Rad. Det. Tech. Meth.}\ }\textbf {\bibinfo
  {volume} {3}},\ \bibinfo {pages} {14} (\bibinfo {year} {2019})}\BibitemShut
  {NoStop}%
\bibitem [{\citenamefont {Li}\ \emph {et~al.}(2017)\citenamefont {Li} \emph
  {et~al.}}]{etof::Li2017}%
  \BibitemOpen
  \bibfield  {author} {\bibinfo {author} {\bibfnamefont {X.}~\bibnamefont {Li}}
  \emph {et~al.},\ }\href@noop {} {\bibfield  {journal} {\bibinfo  {journal}
  {Rad. Det. Tech. Meth.}\ }\textbf {\bibinfo {volume} {1}},\ \bibinfo {pages}
  {13} (\bibinfo {year} {2017})}\BibitemShut {NoStop}%
\bibitem [{\citenamefont {Ablikim}\ \emph {et~al.}(2022)\citenamefont {Ablikim}
  \emph {et~al.}}]{nJpsi}%
  \BibitemOpen
  \bibfield  {author} {\bibinfo {author} {\bibfnamefont {M.}~\bibnamefont
  {Ablikim}} \emph {et~al.} (\bibinfo {collaboration} {BESIII Collaboration}),\
  }\href@noop {} {\bibfield  {journal} {\bibinfo  {journal} {Chin. Phys. C}\
  }\textbf {\bibinfo {volume} {46}},\ \bibinfo {pages} {074001} (\bibinfo
  {year} {2022})}\BibitemShut {NoStop}%
\bibitem [{\citenamefont {Jadach}\ \emph {et~al.}(2000)\citenamefont {Jadach}
  \emph {et~al.}}]{ref::kkmc}%
  \BibitemOpen
  \bibfield  {author} {\bibinfo {author} {\bibfnamefont {S.}~\bibnamefont
  {Jadach}} \emph {et~al.},\ }\href@noop {} {\bibfield  {journal} {\bibinfo
  {journal} {{Comp. Phys. Comm.}}\ }\textbf {\bibinfo {volume} {130}},\
  \bibinfo {pages} {260 } (\bibinfo {year} {2000})}\BibitemShut {NoStop}%
\bibitem [{\citenamefont {Lange}(2001)}]{ref::evtgen2001}%
  \BibitemOpen
  \bibfield  {author} {\bibinfo {author} {\bibfnamefont {D.~J.}\ \bibnamefont
  {Lange}},\ }\href {https://doi.org/10.1016/S0168-9002(01)00089-4} {\bibfield
  {journal} {\bibinfo  {journal} {Nucl. Instr. Meth. A}\ }\textbf {\bibinfo
  {volume} {462}},\ \bibinfo {pages} {152} (\bibinfo {year}
  {2001})}\BibitemShut {NoStop}%
\bibitem [{\citenamefont {Ping}(2008)}]{ref::evtgen2008}%
  \BibitemOpen
  \bibfield  {author} {\bibinfo {author} {\bibfnamefont {R.-G.}\ \bibnamefont
  {Ping}},\ }\href@noop {} {\bibfield  {journal} {\bibinfo  {journal} {Chin.
  Phys. C}\ }\textbf {\bibinfo {volume} {32}},\ \bibinfo {pages} {599}
  (\bibinfo {year} {2008})}\BibitemShut {NoStop}%
\bibitem [{\citenamefont {Agostinelli}\ \emph {et~al.}(2003)\citenamefont
  {Agostinelli} \emph {et~al.}}]{Agostinelli:2002hh}%
  \BibitemOpen
  \bibfield  {author} {\bibinfo {author} {\bibfnamefont {S.}~\bibnamefont
  {Agostinelli}} \emph {et~al.} (\bibinfo {collaboration} {GEANT4}),\
  }\href@noop {} {\bibfield  {journal} {\bibinfo  {journal} {Nucl. Instr. Meth.
  A}\ }\textbf {\bibinfo {volume} {506}},\ \bibinfo {pages} {250} (\bibinfo
  {year} {2003})}\BibitemShut {NoStop}%
\bibitem [{\citenamefont {Chen}\ \emph {et~al.}(2000)\citenamefont {Chen},
  \citenamefont {Huang}, \citenamefont {Qi}, \citenamefont {Zhang},\ and\
  \citenamefont {Zhu}}]{ref:lundcharm1}%
  \BibitemOpen
  \bibfield  {author} {\bibinfo {author} {\bibfnamefont {J.~C.}\ \bibnamefont
  {Chen}}, \bibinfo {author} {\bibfnamefont {G.~S.}\ \bibnamefont {Huang}},
  \bibinfo {author} {\bibfnamefont {X.~R.}\ \bibnamefont {Qi}}, \bibinfo
  {author} {\bibfnamefont {D.~H.}\ \bibnamefont {Zhang}},\ and\ \bibinfo
  {author} {\bibfnamefont {Y.~S.}\ \bibnamefont {Zhu}},\ }\href@noop {}
  {\bibfield  {journal} {\bibinfo  {journal} {Phys. Rev. D}\ }\textbf {\bibinfo
  {volume} {62}},\ \bibinfo {pages} {034003} (\bibinfo {year}
  {2000})}\BibitemShut {NoStop}%
\bibitem [{\citenamefont {Yang}\ \emph {et~al.}(2014)\citenamefont {Yang} \emph
  {et~al.}}]{ref:lundcharm2}%
  \BibitemOpen
  \bibfield  {author} {\bibinfo {author} {\bibfnamefont {R.~L.}\ \bibnamefont
  {Yang}} \emph {et~al.},\ }\href@noop {} {\bibfield  {journal} {\bibinfo
  {journal} {Chin. Phys. Lett.}\ }\textbf {\bibinfo {volume} {31}},\ \bibinfo
  {pages} {061301} (\bibinfo {year} {2014})}\BibitemShut {NoStop}%
\bibitem [{\citenamefont {Fritsch}\ \emph {et~al.}()\citenamefont {Fritsch}
  \emph {et~al.}}]{ComPWA}%
  \BibitemOpen
  \bibfield  {author} {\bibinfo {author} {\bibfnamefont {M.}~\bibnamefont
  {Fritsch}} \emph {et~al.} (\bibinfo {collaboration} {ComPWA}),\ }\href@noop
  {} {\bibinfo  {journal} {DOI: 10.5281/zenodo.5526360, 10.5281/zenodo.5526648,
  10.5281/zenodo.5526650}\ }\BibitemShut {NoStop}%
\bibitem [{\citenamefont {Barlow}(2002)}]{barlow}%
  \BibitemOpen
\bibfield  {journal} {  }\bibfield  {author} {\bibinfo {author} {\bibfnamefont
  {R.}~\bibnamefont {Barlow}},\ }\href@noop {} {\bibfield  {journal} {\bibinfo
  {journal} {Conference on Advanced Statistical 37 Techniques in Particle
  Physics}\ ,\ \bibinfo {pages} {134}} (\bibinfo {year} {2002})}\BibitemShut
  {NoStop}%
\bibitem [{\citenamefont {Ablikim}\ \emph {et~al.}(2014)\citenamefont {Ablikim}
  \emph {et~al.}}]{combination_formula}%
  \BibitemOpen
  \bibfield  {author} {\bibinfo {author} {\bibfnamefont {M.}~\bibnamefont
  {Ablikim}} \emph {et~al.} (\bibinfo {collaboration} {BESIII Collaboration}),\
  }\href@noop {} {\bibfield  {journal} {\bibinfo  {journal} {Phys. Rev. D}\
  }\textbf {\bibinfo {volume} {89}},\ \bibinfo {pages} {074030} (\bibinfo
  {year} {2014})}\BibitemShut {NoStop}%
\end{thebibliography}%

\end{document}